# An NLP-Driven Approach Using Twitter Data for Tailored K-pop Artist Recommendations

Sora Kang[1], Mingu Lee[2*]
[1]Seoul National University, Seoul, Korea
[2]Seoul National University, Seoul, Korea
E-mail: sorakang@snu.ac.kr; hybop@snu.ac.kr

*Abstract*— **The global rise of K-pop and the digital revolution have paved the way for new dimensions in artist recommendations. With platforms like Twitter serving as a hub for fans to interact, share and discuss K-pop, a vast amount of data is generated that can be analyzed to understand listener preferences. However, current recommendation systems often overlook K-pop's inherent diversity, treating it as a singular entity. This paper presents an innovative method that utilizes Natural Language Processing to analyze tweet content and discern individual listening habits and preferences. The mass of Twitter data is methodically categorized using fan clusters, facilitating granular and personalized artist recommendations. Our approach marries the advanced GPT-4 model with large-scale social media data, offering potential enhancements in accuracy for K-pop recommendation systems and promising an elevated, personalized fan experience. In conclusion, acknowledging the heterogeneity within fanbases and capitalizing on readily available social media data marks a significant stride towards advancing personalized music recommendation systems.**

## I. INTRODUCTION

Korean pop music, better known as K-pop, has made monumental strides in its global popularity over the past decade [5,8]. The worldwide fan engagement is evident from the stunning 7.8 billion global Tweets related to K-pop in 2021, surpassing the previous record of 6.7 billion Tweets in 2020 by a considerable 16% [9]. Such an increase underscores K-pop's expanding reach and the vibrant, diverse conversations springing forth within its fandom on social media platforms such as Twitter.

Twitter, a platform frequently used by fans to express their connection to, and affection for, their favorite K-pop artists and sub-genres, serves as a valuable source of data for discerning music preferences and listening habits [2,3,10]. This vast data, unfortunately, often remains underutilized even as it holds the key to the customization and personalization craved by users in their music experiences.

Music recommendation systems have emerged as an integral element shaping the user experience on digital music platforms [7,11]. These automated systems, ranging from collaborative filtering methods, and content-based methods to hybrid methods, have been the focus of intense research [1,12]. They mold user satisfaction by providing recommendations tailor-made to individual tastes and preferences. In response to keeping up with the burgeoning K-pop market, this paper aims to provide a more fit-for-purpose recommendation system using GPT-4 model to analyze Tweet content and create distinct clusters of K-pop fans based on their diverse interests and behaviors. This innovative approach promises to enhance the specificity and usefulness of K-pop artist recommendations, thereby elevating the overall user experience.

## II. METHODS

To enhance the comprehensiveness of our discourse analysis, we meticulously compiled a rich and varied dataset of tweets related to the K-pop genre. We conducted a systematic crawl and aggregation of tweets that mentioned a noteworthy 263 active K-pop groups from a comprehensive list of 426 groups (see Table 1). This careful selection was aimed at encompassing a wide spectrum of discussions, perspectives, and fan preferences within the K-pop community on the prominent social networking platform, Twitter. Over the span of approximately two weeks, from October 10 to October 24, 2023, our data collection efforts yielded a corpus of approximately 640K tweets.

Table 1 Sample of K-pop Groups Included in Study

| Name | Gender | Debut | Agency | Size | Active |
|---|---|---|---|---|---|
| **(G)I-DLE** | Female | 2/05/18 | Cube | 5 | Yes |
| **KARA** | Female | 29/03/07 | DSP | 5 | No |
| **14U** | Male | 17/04/17 | BG | 14 | No |
| **15&** | Female | 5/10/12 | JYP | 2 | No |
| **1TEAM** | Male | 27/03/19 | Liveworks | 5 | No |
| **iKON** | Male | 15/09/15 | YG | 6 | Yes |

We then preprocessed this data by eliminating irrelevant components including URLs, hashtags, and user mentions, whilst normalizing the text through case folding and tokenization. Text clustering techniques backed by TF-IDF vectorizer and K-means clustering algorithm were employed to transform this corpus into a numerical matrix and partition the same into nine distinct fan categories. These categories identified were "Vocal Talent Admirers", "Merchandise Buyers", "Content Creators", "Concert-Goers", "Information Spreaders", "Activists", "Debaters", and "Language Learners".

Figure 1 Word Cloud of K-pop related Tweets

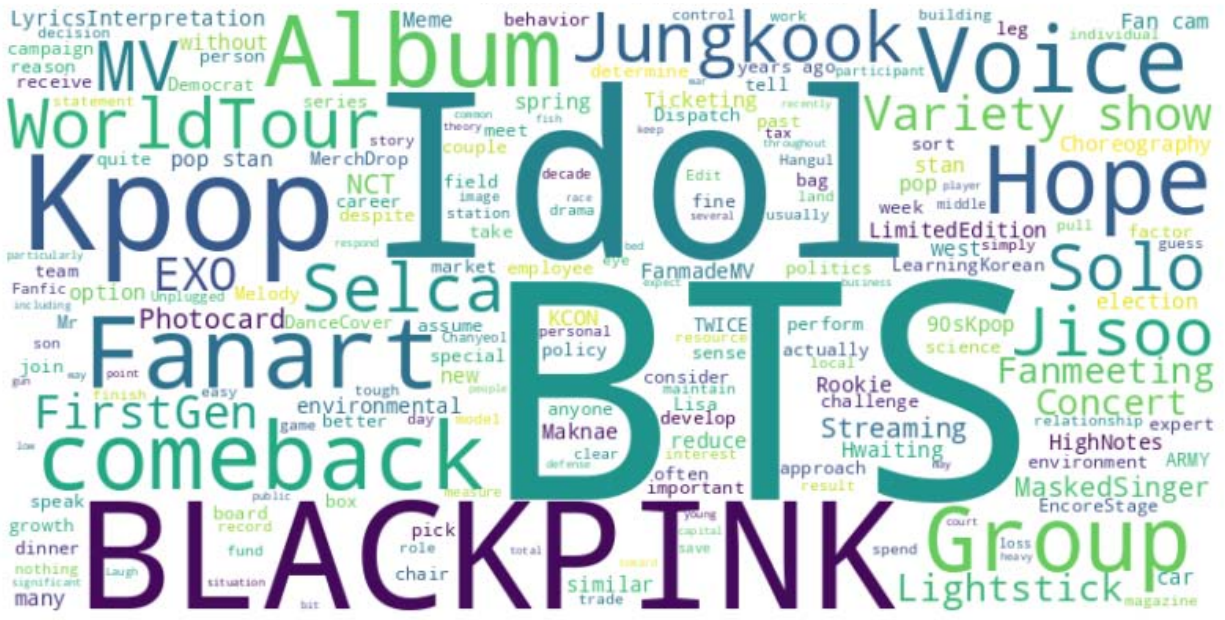

Further refining our analysis, we refined our fan categories to focus solely on those directly related to musical preferences, excluding "Information Spreaders", "Activists", and "Debaters". With the refined clusters, we utilized a Generative Pretrained Transformer 4 (GPT-4) model to generate accurate music recommendations. The GPT-4, the latest in language processing AI technology, was chosen for its impressive ability to understand and generate contextually nuanced human-like text based on an extensive training dataset[6]. As the GPT-4 model has been trained on diverse data sources, including past tweets, album reviews, and fan forums, it was able to cognitively draw relationships between different K-pop artists and fan preferences[4]. By inputting the characterized fan clusters into this model, we created a list of recommended K-pop artists tailored for each cluster.

## III. RESULTS

The refined fan clusters and the use of GPT-4 created the ability to generate specific musical recommendations catering to delineated interests within the fandom. The following table summarizes the final results of our analysis. Only selected top keywords and artists for each fan cluster are included in Table 2 to present a succinct summary:

Table 2 Recommended Artists based on K-pop Fandom Cluster

| Cluster | Description | Keywords | Artists |
|---|---|---|---|
| **Vocal Talent Admirers** | Fans who highly appreciate the vocal talents of K-pop idols. | HighNotes, Voice, MaskedSinger, Unplugged | Mamamoo, Ailee, EXO, Gummy, Kim Feel |
| **Merchandise Buyers** | Fans who invest in buying K-pop merchandise. | MerchDrop, LimitedEdition, Album, Photocard | Seventeen, NCT, ATEEZ, CosmicGirls, Loona |
| **Content Creators** | Fans who create K-pop-related content like fan art. | Fanart, Edit, Fanfic, DanceCover, FanmadeMV | Stray Kids, ATEEZ, MONSTA X, GFRIEND |
| **Concert-Goers** | Fans who attend K-pop concerts and discuss live performances. | Ticketing, WorldTour, EncoreStage | BIGBANG, BTS, EXO, Mamamoo, GOT7, PSY |
| **Retro Music Fans** | Fans who appreciate older K-pop music of '90s to 2000s. | 90sKpop, years ago, comeback FirstGen, | H.O.T., Sechs Kies, TVXQ, BoA, S.E.S, Rain |
| **Language Learners** | Fans interested in learning Korean through K-pop. | LearningKorean, Hangul, LyricsInterpretation | Epik High, Heize, ZICO, AKMU, Gaho, Baek Yerin |

Through these results, our study achieved its aim of identifying unique K-pop fan categories on Twitter and linking them with suitable K-pop artist recommendations. The insights gained from this work bear significant implications for music recommendation systems and the broader music industry. By tapping into the wealth of social media data, we can better understand and cater to fan preferences, thereby promoting a more personalized, fan-centered music experience. With further research and refinement, this approach has the potential to revolutionize the way we engage with music fandoms.

## IV. CONCLUSION

This study offers an innovative approach to enhancing K-pop artist recommendations using Twitter data and natural language processing through the use of the GPT-4 model. We've demonstrated the vast diversity within the K-pop fandom on Twitter, challenging conventional broad-based fan categorizations and unveiling a range of fan interests and engagement styles.

Our research illustrates the potential of AI in music recommendation, particularly highlighting the effectiveness of the GPT-4 model in leveraging social media data to generate personalized music suggestions. Despite limitations such as the focus on English language tweets and Twitter as the sole platform, this study sets a precedent for the use of AI and social media analysis in music fandom understanding and recommendation system enhancement.

In conclusion, our study spotlighted unique insights into fan categorizations and music recommendations using a data-driven approach with LLM, painting a rich picture of the K-pop fandom. The outcomes of this research encourage further exploration in this field, potentially guiding the future of personalized music recommendation systems.

## ACKNOWLEDGEMENT

This research was supported by the AIIS(Artificial Intelligence Institute, Seoul National University), under the support program(0670-20230065) supervised by the AI applied design course from Seoul National University.